\documentclass[prd,nofootinbib]{revtex4}
\usepackage{graphicx}
\usepackage{amsmath}
\usepackage{amssymb}
\usepackage{float}
\usepackage{color}

\usepackage{latexsym}
\usepackage{array}
\usepackage{epic}
\usepackage{eepic}
\usepackage{cancel}
\usepackage[mathscr]{euscript}

\begin{document}
\title{Dirac and non-Dirac conditions in the 2-potential theory of magnetic charge}

\author{John Scott}
\email{johnmlscott@mail.fresnostate.edu}
\affiliation{Department of Physics, California State University Fresno, Fresno, CA 93740-8031, USA}

\author{Douglas Singleton}
\email{dougs@mail.fresnostate.edu}
\affiliation{Department of Physics, California State University Fresno, Fresno, CA 93740-8031, USA \\
and \\
Institute of Experimental and Theoretical Physics Al-Farabi KazNU, Almaty, 050040, Kazakhstan}

\author{Timothy J. Evans}
\email{tevans559@mail.fresnostate.edu}
\affiliation{Department of Physics, California State University Fresno, Fresno, CA 93740-8031, USA}

\author{
Vladimir Dzhunushaliev
}
\email{v.dzhunushaliev@gmail.com}
\affiliation{
Dept. Theor. and Nucl. Phys., KazNU, Almaty, 050040, Kazakhstan
}
\affiliation{
IETP, Al-Farabi KazNU, Almaty, 050040, Kazakhstan
}

\author{Vladimir Folomeev}
\email{vfolomeev@mail.ru}
\affiliation{Institute of Physicotechnical Problems and Material Science of the NAS
of the Kyrgyz Republic, 265 a, Chui Street, Bishkek 720071,  Kyrgyz Republic}

\date{\today}

\begin{abstract}
We investigate the Cabbibo-Ferrari, two potential approach to magnetic charge coupled to two different complex scalar fields, $\Phi_1$ and 
$\Phi_2$, each having different electric and magnetic charges. The scalar field, $\Phi_1$, is assumed to have a spontaneous symmetry breaking self interaction potential which gives a mass to the ``magnetic" gauge potential and  ``magnetic" photon, while the other ``electric" gauge potential and ``electric" photon remain massless. The magnetic photon is hidden until one reaches energies of the order of the magnetic photon rest mass. The second scalar field, $\Phi _2$, is required in order to make the theory non-trivial. With only one field one can always use a duality rotation to rotate away either the electric or magnetic charge, and thus decouple either the associated electric or magnetic photon. In analyzing this system of two scalar fields in the Cabbibo-Ferrari approach we perform several duality and gauge transformations, which require introducing non-Dirac conditions on the initial electric and magnetic charges. We also find that due to the symmetry breaking the usual Dirac condition is altered to include the mass of the magnetic photon. We discuss the implications of these various conditions on the charges.
\end{abstract}
\maketitle

\section{Introduction}
Maxwell's equations with magnetic sources in Gaussian units are
\begin{equation}
\label{maxwellequations}
\begin{split}
&\nabla\cdot{\bf E} = 4 \pi \rho_e \ \ \ \ \ -\nabla\times{\bf E}=\frac{4 \pi}{c}\left(\frac{\partial{\bf B}}{\partial t}+{\bf J}_m\right)\\
&\nabla\cdot{\bf B} = 4 \pi \rho_m \ \ \ \ \ \ \ \ \nabla\times{\bf B}=\frac{4 \pi}{c}\left(\frac{\partial{\bf E}}{\partial t}+{\bf J}_e\right) ~.
\end{split}
\end{equation}
Eqns. \eqref{maxwellequations} are more symmetric than Maxwell equations with only electric sources. The mathematical expression of this extra symmetry is given by the dual transformation of electric and magnetic quantities \cite{jackson}
\begin{equation}
\label{duality}
{\rm Electric} = \cos (\theta ) ~{\rm Electric}' + \sin (\theta) ~ {\rm Magnetic}' \ \ \ \ \ {\rm Magnetic} = - \sin (\theta )~ {\rm Electric}' + \cos (\theta ) ~ {\rm Magnetic}' ~,
\end{equation}
where ${\rm Electric}'$ (${\rm Magnetic '}$) are the original quantities: ${\bf E}'$, $\rho _e '$, ${\bf J}_e '$ (${\bf B}'$, $\rho _m'$, ${\bf J}_m '$) and ${\rm Electric}$ (${\rm Magnetic}$) are the final quantities: ${\bf E}$, $\rho _e$, ${\bf J}_e $ (${\bf B} $, $\rho _m $, ${\bf J}_m $), and $\theta$ is the duality angle by which these quantities are rotated in ``dual" space. This dual symmetry does not usually extend down the level of the potentials, since the electric and magnetic fields in terms of the scalar potential, $\phi_e$, and vector potential, ${\bf A}$, take the asymmetric form ${\bf E}=-\nabla\phi_{e}-\frac{1}{c}\frac{\partial{\bf A}}{\partial t}$ and ${\bf B}= \nabla\times {\bf A}$. In order to extend the dual symmetry of \eqref{duality} to the level of the potentials one needs to introduce
a second set of potentials -- a strategy first proposed by Cabibbo and Ferrari \cite{cabibbo} who added a second scalar and 
vector potential, $\phi_m$ and ${\bf C}$. In terms of these two potentials the electric and magnetic fields can be written as 
\begin{equation}
\label{newEB-pot}
{\bf E}=-\nabla\phi_{e}-\frac{1}{c}\frac{\partial{\bf A}}{\partial t}-\nabla\times{\bf C} ~~~~;~~~~
{\bf B}=-\nabla\phi_{m}-\frac{1}{c}\frac{\partial{\bf C}}{\partial t} + \nabla\times{\bf A} ~.
\end{equation}
With the definitions in \eqref{newEB-pot} the dual symmetry of \eqref{duality} can be extended to the level of the potentials with ${\rm Electric} / {\rm Magnetic}$ in \eqref{duality} being $(\phi_e , {\bf A}) / (\phi_m , {\bf C})$. The 2-potential approach avoids the Dirac string \cite{dirac} which is the usual way magnetic charge is introduced with only a single 4-vector potential
(see for example \cite{ryder} pgs. 402-405 for a short pedagogical exposition of the Dirac string approach). A short, non-exhaustive list of work on the Cabibbo-Ferrari approach to magnetic charge can be found in references \cite{hagen, salam, zwang, barker, singleton1, singleton2,singleton3,chatterjee,chatterjee2}. 

From equation \eqref{newEB-pot} one can work out the behavior of $\phi_m$ and ${\bf C}$ under the discrete $P$ and $T$ symmetries ({\it i.e.} parity or space inversion and time reversal). Given that ${\bf E}$ is odd under $P$ and even under $T$, while ${\bf B}$ is the opposite, one finds from equation \eqref{newEB-pot} that $\phi_m$ is a pseudoscalar (odd under $P$ and odd under $T$) and ${\bf C}$ is a pseudovector (even under $P$ and even under $T$). 

While the Cabibbo-Ferrari approach avoids the need for a singular Dirac string this comes at the expense of introducing extra degrees of freedom via $\phi_m$ and ${\bf C}$. Because of the opposite behavior of $\phi_m$ and ${\bf C}$ under parity and time reversal as compared to $\phi_e$ and ${\bf A}$ these ``magnetic" potentials would give rise to a pseudo-vector, massless ``magnetic" photon in the quantized version of the theory. There are various ways of dealing with these extra degrees of freedom: (i) One can impose extra conditions on the theory so that despite having two 4-vector potentials the final number of degrees of freedom correspond to only a single ``electric" photon; (ii)  accept the second 4-potential as a real ``magnetic" photon \cite{salam} but then apply the Higgs mechanism so that this additional photon becomes massive and is ``hidden" until one reaches some appropriately, high energy scale \cite{singleton1, singleton2, singleton3}. It is this latter choice that we have in mind here. This then expands the usual $SU(2) \times U(1)$ Standard Model to $SU(2) \times U(1) \times U(1)$. Introducing an extra, massive $U(1)$ gauge boson may be phenomenologically interesting in light of dark/hidden photon models \cite{darkphoton} as well as light-shining-through-wall experiments \cite{wall}, all of which require an extra $U(1)$ gauge boson. The magnetic photon might provide an electromagnetic origin for these dark/hidden photon models. In addition given the opposite behavior of $A_\mu$ and $C_\mu$ under $P$ and $T$, as discussed above, the ``electric" photon associated with $A_\mu$ and the ``magnetic" photon associated with $C_\mu$ will different properties under $P$ and $T$ symmetry and will thus be distinguishable on this basis.   

The above results can be put in 4-vector notation. The electric and magnetic 4-vector  potentials are $A ^ \mu =(\phi_e, {\bf A})$ and $C ^ \mu =(\phi_e, {\bf A})$, and the electric and magnetic 4-currents are $J^{\mu}_{e}=\left(\rho_{e},{\bf J}_{e}\right)$ and $J^{\mu}_{m}=\left(\rho_{m},{\bf J}_{m}\right)$. From the electric and magnetic 4-vector  potentials one can construct the electric and magnetic field strength tensors and their dual field strength tensors
\begin{eqnarray}
\label{fsEB}
F^{\mu\nu} &=& \partial^{\mu}A^{\nu}-\partial^{\nu}A^{\mu}  ~~~~ ; ~~~~G^{\mu\nu}=\partial^{\mu}C^{\nu}-\partial^{\nu}C^{\mu} \nonumber \\
{\cal F}^{\mu \nu} &=& \frac{1}{2} \epsilon ^{\mu \nu \alpha \beta} F_{\alpha \beta} ~~~; ~~~
{\cal G}^{\mu \nu}  = \frac{1}{2} \epsilon ^{\mu \nu \alpha \beta} G_{\alpha \beta}~,
\end{eqnarray}
In terms of these the electric and magnetic fields from \eqref{newEB-pot} can be written as $E^i = F^{i0} - {\cal G}^{i0}$ and $B^i = G^{i0} + {\cal F}^{i0}$. Using all this one can construct a pure electromagnetic field Lagrangian density
\begin{equation}
\label{lagrane-em}
{\cal L} = -\frac{1}{16 \pi} F_{\mu \nu} F^{\mu \nu} -\frac{1}{16 \pi} G_{\mu \nu} G^{\mu \nu} 
+ K_1 \epsilon ^{\mu \nu \alpha \beta}F_{\mu \nu} G_{\alpha \beta} + K_2 F_{\mu \nu} G^{\mu \nu} -J^\mu _e A_\mu - J^\mu _m C_\mu ~,
\end{equation}
where for generality we have included two additional mixing terms between the two different field strength tensors -- 
$K_1 \epsilon ^{\mu \nu \alpha \beta}F_{\mu \nu} G_{\alpha \beta}$ and $K_2 F_{\mu \nu} G^{\mu \nu}$ -- with $K_1, K_2$ being undetermined constants. The $K_1$ term does not effect the equations of motions. The $K_2$ term would change the equations of motion and is a kinetic mixing term of the kind found in many dark/hidden photon models \cite{darkphoton} and light-shining-through-wall \cite{wall} proposals. Neither term will play a role in the considerations in the rest of this paper and so we will drop both terms in the following sections. With the dropping of the $K_2$ term the equations of motion resulting from \eqref{lagrane-em} are $\partial _\mu F^{\mu \nu} = 4 \pi J^\nu _e$ and $\partial _\mu G^{\mu \nu} = 4 \pi J^\nu _m$ which are the 4-vector form of the equations in \eqref{maxwellequations}. (Here and in the rest of the article we set $c=1$). In the next section we replace the arbitrary sources $J^\mu _e , J^\mu _m$ by a dynamical source coming from a complex scalar field initially carrying both electric and magnetic charge.

\section{Scalar fields coupled to electric and magnetic charge}   

In this section we couple the two, 4-vector potentials, $A_\mu$ and $C_\mu$, to two different complex scalar field, $\Phi _1 , \Phi_2$. We do this in steps -- first coupling $A_\mu , C_\mu$ to $\Phi_1$ (in section IIA) and then coupling $A_\mu , C_\mu$ to $\Phi_2$ (in section IIB). The scalar field $\Phi_1$ has a $\lambda \Phi_1 ^4$ self-interaction which spontaneously breaks the magnetic gauge symmetry of the theory, making the magnetic photon massive. This makes the magnetic photon unobservable until one probes an energy scale on the order of the rest mass energy of the magnetic photon. However with only a single scalar field $\Phi _1$ we find after symmetry breaking that the electric 4-vector potential, $A_\mu$, completely decouples from $\Phi _1$ and the theory becomes trivial in regard to $A_\mu$. This triviality goes back to the well known result that if all fields/particles have the same ratio of electric charge to magnetic charge (which is true with only one scalar field) one can always perform a duality rotation to rotate away either the electric or magnetic charge. To make the theory non-trivial we introduce a second complex scalar field $\Phi_2$ with different initial charges from $\Phi_1$.   

\addcontentsline{toc}{section}{Scalar Fields Lagrangian coupled to electric and magnetic charge}   

\subsection{Scalar field $\Phi_1$}

We begin with one complex, scalar field, $\Phi_1$, with a symmetry breaking self-interacting potential. $\Phi _1$ initially couples to both $A_\mu$ and $C_\mu$ with strength $q_e$ and $q_m$ respectively.  $\Phi_1$ provides a dynamical source for the 4-currents $J^\mu _e$ and $J^\mu _m$. The following manipulations are the standard symmetry breaking of a $U(1) \times U(1)$ gauge symmetry down to $U(1)$. We repeat the details here since it will serve as necessary background for the following section IIB. The Lagrangian density is
\begin{equation}
\label{Lscalar}
{\cal L}_{1}= D_{\mu}\Phi_1^*D^{\mu}\Phi_1-\frac{1}{16 \pi}F_{\mu\nu}F^{\mu\nu}-\frac{1}{16 \pi}G_{\mu\nu}G^{\mu\nu} - V(\Phi _1) ~,
\end{equation}
where the gauge-covariant derivative is $D^{\mu}=\partial^{\mu}-iq_e A^{\mu}-iq_m C^{\mu}$. There are two gauge field kinetic terms, $-\frac{1}{16 \pi}F_{\mu\nu}F^{\mu\nu}$ and $-\frac{1}{16 \pi}G_{\mu\nu}G^{\mu\nu}$, for $A_\mu$ and $C_\mu$ respectively. The self interaction term for $\Phi_1$ is of the form $V(\Phi _1) = m^2 \Phi_1^* \Phi_1 + \lambda (\Phi_1^* \Phi_1)^2$ and the mass term is tachyonic ({\it i.e.} $m^2 <0$). Under these conditions $\Phi_1$ develops a vacuum expectation value
\begin{equation}
\label{vev}
|\Phi _1|=\sqrt {(\Phi_1 ^* ~ \Phi_1)} = \left(\frac{-m^2}{2\lambda}\right)^{1/2}\equiv \frac{v}{\sqrt{2}} ~,
\end{equation}
where $v= \sqrt{-m^2/\lambda}$. Expanding $\Phi_1$ around $v$ gives
\begin{equation}
\label{vev2}
\Phi _1 = \frac{1}{\sqrt{2}}\left(v+\eta(x)+i\zeta(x)\right) \approx \frac{1}{\sqrt{2}}\left(v+\eta(x)\right)e^{i\zeta(x)/v}~,
\end{equation}
where $\eta (x) , \zeta (x)$ are real fields. Following the standard prescription we perform a gauge transformation to get rid of the phase, $e^{i\zeta(x)/v}$, in $\Phi _1$
\begin{eqnarray}
\label{phiapprox}
\Phi_1'(x)&=&e^{-i\zeta(x)/v}\Phi_1 (x)\approx \frac{1}{\sqrt{2}}\left(v+\eta(x)\right) ~, \nonumber \\
A'_{\mu}(x)&=&A_{\mu}(x)-\frac{1}{2q_e v}\partial_{\mu}\zeta(x) ~, \\
C'_{\mu}(x)&=&C_{\mu}-\frac{1}{2q_m v}\partial_{\mu}\zeta(x) ~.\nonumber
\end{eqnarray}
The Lagrangian in \eqref{Lscalar} now becomes
\begin{eqnarray}
\label{Lscalar2}
{\cal L}_{1} &=& \frac{1}{2} \partial_{\mu} \eta \partial^{\mu} \eta - \frac{1}{16 \pi}F_{\mu\nu}F^{\mu\nu}-\frac{1}{16 \pi}G_{\mu\nu}G^{\mu\nu}
+ m^2\eta^2(x) - \lambda v\eta^3(x) - \frac{1}{4}\lambda\eta^4(x) \\
&+& \frac{1}{2}\left(v^2+2v\eta(x)+\eta^2(x)\right) (q_e^2A_{\mu}A^{\mu}+q_m^2C_{\mu}C^{\mu}+2q_e q_mA_{\mu}C^{\mu} ) ~. \nonumber
\end{eqnarray}
In \eqref{Lscalar2} we have dropped a constant term, $m^2 v^2 /2$, which has no effect on the equations of motion. Also we have relabeled the gauge fields, by dropped the primes. The second line in \eqref{Lscalar2} represents 
a mass term for the scalar field $\eta$ as well as interaction terms between $\eta$ and {\it both} $A_\mu$ and $C_\mu$. The mixed term 
$2q_e q_m A_{\mu}C^{\mu}$ is problematic. We can get rid of this term by performing a duality rotation 
on $A_\mu$ and $C_\mu$. This is exactly analogous to the rotation by the Weinberg angle that one performs in the Standard Model in order to 
get the physical Z-boson and photon fields from the initial $W^0_\mu$ and $X_\mu$ bosons. Writing the second line in \eqref{Lscalar2} as 
\begin{eqnarray}
\label{AC-inter}
\left(\frac{1}{2}v^2+v\eta(x)+\frac{1}{2}\eta^2(x)\right) \left(A_{\mu}, C_{\mu} \right)\left(
                                    \begin{array}{cc}
                                      q_e^2 & q_eq_m \\
                                      q_eq_m & q_m^2 \\
                                    \end{array}
                                  \right)\left(
                                           \begin{array}{c}
                                             A^{\mu} \\
                                             C^{\mu} \\
                                           \end{array}
                                         \right) ~,
\end{eqnarray}
we perform the following duality rotation of the potentials 
\begin{equation}
\label{AC-dual}
A^{\mu } \rightarrow A^{\mu \prime}\cos\theta+C^{\mu \prime}\sin\theta ~~~~~~ ; ~~~~~~~
C^{\mu } \rightarrow -A^{\mu \prime}\sin\theta +C^{\mu \prime}\cos\theta ~.
\end{equation}
This transforms the gauge field part of \eqref{AC-inter} into 
\begin{eqnarray}
\label{AC-inter2}
&&\left(A_{\mu}^{\prime}, C_{\mu}^{\prime}\right)\left(
                                           \begin{array}{cc}
                                             \cos\theta & -\sin\theta \\
                                             \sin\theta & \cos\theta \\
                                           \end{array}
                                         \right)\left(
                                    \begin{array}{cc}
                                      q_e^2 & q_eq_m \\
                                      q_eq_m & q_m^2 \\
                                    \end{array}
                                  \right)\left(
                                           \begin{array}{cc}
                                             \cos\theta & \sin\theta \\
                                             -\sin\theta & \cos\theta \\
                                           \end{array}
                                         \right)
                                  \left(
                                           \begin{array}{c}
                                             A^{\mu \prime} \\
                                             C^{\mu \prime} \\
                                           \end{array}
                                         \right) ~.
\end{eqnarray}
To diagonalize the gauge field interaction matrix of \eqref{AC-inter2} one can choose $\theta$ so that $\cos\theta = \frac{q_m}{\sqrt{q_e^2+q_m^2}}$ and $\sin\theta = \frac{q_e}{\sqrt{q_e^2+q_m^2}}$. 
In this way the interaction matrix of \eqref{AC-inter2} diagonalizes to become 
\begin{equation}
\label{AC-inter2a}
\left(\frac{1}{2}v^2+\frac{1}{2}v\eta(x) + \frac{1}{2}\eta^2(x)\right) \left(A_{\mu}^{\prime}, C_{\mu}^{\prime}\right)\left(
                                           \begin{array}{cc}
                                             0 & 0 \\
                                             0 & q_e^2+q_m^2 \\
                                           \end{array}
                                         \right)
                                  \left(
                                           \begin{array}{c}
                                             A^{\mu \prime} \\
                                             C^{\mu \prime} \\
                                           \end{array}
                                         \right) ~.
\end{equation}
This can be written out as
\begin{equation}
\label{CC-inter}
\left(\frac{1}{2}v^2+\frac{1}{2}v\eta(x) + \frac{1}{2}\eta^2(x)\right)\left(q_e^2+q_m^2\right)C_{\mu}^{\prime}C^{\mu \prime} ~.
\end{equation}
Piecing together all of these parts we find that the Lagrangian from \eqref{Lscalar2} becomes,
\begin{eqnarray}
\label{Lscalar3}
{\cal L}_1 &=& \frac{1}{2}\partial_{\mu}\eta(x)\partial^{\mu}\eta(x)+\frac{1}{2}(2m^2)\eta^2(x)-\frac{1}{16 \pi}F_{\mu\nu}F^{\mu\nu}-\frac{1}{16 \pi}G_{\mu\nu}G^{\mu\nu} \nonumber \\
&+& \left(\frac{1}{2}v^2+\frac{1}{2}v\eta(x) + \frac{1}{2}\eta^2(x)\right)\left(q_e^2+q_m^2\right)C_{\mu}C^{\mu}-\lambda v\eta^3(x)-\frac{1}{4}\lambda\eta^4(x).
\end{eqnarray}
In the above we have again dropped the primes. Note that the electric gauge potential, $A^\mu$, is completely
decoupled from the scalar matter field, $\eta$, and from $C_\mu$ ; $A_\mu$ has become a sterile/decoupled field in ${\cal L}_1$ since it interacts with nothing else. In the Lagrangian in \eqref{Lscalar3} there is only one interacting gauge potential, $C_\mu$, and only one magnetically charged scalar field, $\eta$. This is related to the fact that in a world with only one type of field/particle one can always rotate away one of the charges via a duality rotation. In summary the final Lagrangian in \eqref{Lscalar3} has a completely decoupled, massless electric gauge potential, $A^\mu$, and a massive ``magnetic" gauge potential, $C^{\mu}$, which is coupled to a scalar field, $\eta$. This scalar field carries only a magnetic charge of $g=\left(q_e^2+q_m^2\right)^{1/2}$. This physical magnetic charge $g$is a combination of the initial charges $q_e$ and $q_m$. The mass term for $C_\mu$ is given by $\frac{1}{2} m_C ^2 C_\mu C^\mu \rightarrow \frac{1}{2} v^2(q_e^2+q_m^2)C_{\mu}C^{\mu}$. Thus the mass associated with $C_\mu$ is $m_C = v \sqrt{q_e^2+q_m^2} = g v$.

\subsection{Scalar field $\Phi_2$}   
\addcontentsline{toc}{section}{Scalar field $\Phi_2$}   

In order to move beyond the trivial results of section IIA, and to obtain the advertised non-Dirac conditions, we 
add a second scalar field, $\Phi_2$, with different initial electric and magnetic charges, $q_e '$ and $q_m '$. 
The Lagrangian from \eqref{Lscalar} now expands to become
\begin{equation}
\label{Lscalar4}
{\cal L}_{1+2}= D_{\mu}\Phi_1^*D^{\mu}\Phi_1+D_{\mu}\Phi_2^*D^{\mu}\Phi_2-\frac{1}{16 \pi}F_{\mu\nu}F^{\mu\nu}-\frac{1}{16 \pi}G_{\mu\nu}G^{\mu\nu}-V(\Phi_1^2) ~.
\end{equation}
The covariant derivatives for the two fields, $\Phi _1$ and $\Phi _2$ are respectively  
$D^{\mu}\Phi_1=\left(\partial^{\mu}-iq_eA^{\mu}-iq_mC^{\mu}\right)\Phi_1$ and $D^{\mu}\Phi_2=\left(\partial^{\mu}-iq_e'A^{\mu}-iq_m'C^{\mu}\right)\Phi_2$. Using these covariant derivatives to expand \eqref{Lscalar4} gives 
\begin{eqnarray}
\label{Lscalar5}
{\cal L}_{1+2} &=& \partial_{\mu}\Phi_1^*\partial^{\mu}\Phi_1+\partial_{\mu}\Phi_2^*\partial^{\mu}\Phi_2+J^{\mu}_1 (q_e A_\mu+ q_m C_\mu)
+J^{\mu}_2 (q'_e A_\mu + q'_m C_\mu ) -\frac{1}{16 \pi}F_{\mu\nu}F^{\mu\nu}-\frac{1}{16 \pi}G_{\mu\nu}G^{\mu\nu} - V(\Phi_1^2) \nonumber \\
&+&\left(q_e ^2 A_{\mu} A^\mu + q_m ^2 C_{\mu} C^\mu + 2 q_e q_m A_\mu C^\mu \right)\Phi_1^*\Phi_1
+\left((q '_e)^2 A_{\mu} A^\mu + (q'_m )^2 C_{\mu} C^\mu + 2 q'_e q'_m A_\mu C^\mu \right)\Phi_2^*\Phi_2 ~,
\end{eqnarray}
where the currents, $J_1 ^\mu$ and $J_2 ^\mu$, from $\Phi_1$ and $\Phi_2$ respectively, are defined as
\begin{equation}
\label{current12}
J^{\mu}_1=i\left[\Phi_1^*\left(\partial^{\mu}\Phi_1\right)-\Phi_1\left(\partial^{\mu}\Phi_1\right)^*\right] ~~~;~~~
J^{\mu}_2=i\left[\Phi_2^*\left(\partial^{\mu}\Phi_2\right)-\Phi_2\left(\partial^{\mu}\Phi_2\right)^*\right]
\end{equation}
As in the previous section, we perform the gauge transformation on $\Phi _1, A_\mu, C_\mu$ given in \eqref{phiapprox}. The $\Phi_1$ part of the ${\cal L}_{1+2}$ reduces as in \eqref{Lscalar2} and the whole Lagrangian from \eqref{Lscalar5} reduces to 
\begin{eqnarray}
\label{Lscalar6}
{\cal L}_{1+2} &=& \frac{1}{2} \partial_{\mu} \eta \partial^{\mu} \eta + \partial_{\mu}\Phi_2^*\partial^{\mu}\Phi_2 - \frac{1}{16 \pi}F_{\mu\nu}F^{\mu\nu}-\frac{1}{16 \pi} G_{\mu\nu}G^{\mu\nu}
+ m^2\eta^2(x) - \lambda v\eta^3(x) - \frac{1}{4}\lambda\eta^4(x) \nonumber \\
&+& \frac{1}{2}\left(v^2+2v\eta(x)+\eta^2(x)\right) (q_e^2A_{\mu}A^{\mu}+q_m^2C_{\mu}C^{\mu}+2q_e q_mA_{\mu}C^{\mu} ) +J^{\mu}_2 (q'_e A_\mu + q'_m C_\mu ) \nonumber \\
&+&\left((q '_e)^2 A_{\mu} A^\mu + (q'_m )^2 C_{\mu} C^\mu + 2 q'_e q'_m A_\mu C^\mu \right)\Phi_2^*\Phi_2 + J^{\mu}_2 \left( \frac{q'_e}{2vq_e} + \frac{q'_m}{2vq_m} \right) \partial_\mu \zeta \\
&+& \frac{1}{4 v^2} \left(\frac{(q '_e)^2}{q_e ^2} + \frac{(q'_m )^2}{q_m ^2} + 2 \frac{q'_e q'_m}{q_e q_m}  \right) (\partial_\mu \zeta) (\partial^\mu \zeta ) \Phi_2^*\Phi_2 \nonumber ~.
\end{eqnarray}
Note that $J_1 ^\mu$ is missing. By the gauge transformation \eqref{phiapprox} $\Phi_1 \rightarrow \frac{1}{\sqrt{2}} \left(v+\eta(x)\right)$ ({\it i.e.} the scalar field is purely real) and so the associated $J_1 ^\mu$ vanishes. The last two terms in \eqref{Lscalar6} depend on the arbitrary gauge function $\zeta (x)$, so the Lagrangian in \eqref{Lscalar6} is not invariant under the
gauge transformation \eqref{phiapprox}. One could try another gauge transformation on $\Phi_2$ in order to restore the invariance of the Lagrangian. However this would then require making an additional gauge transformation of $A_\mu , C_\mu$, which then would require another gauge transformation on $\Phi _1$ {\it etc.} There is another option: require the charges on $\Phi _1$ and $\Phi_2$ satisfy the relationship 
\begin{equation}
\label{gauge-cond}
\frac{q_e'}{q_e}=-\frac{q_m'}{q_m} ~.
\end{equation}
This is the first non-Dirac condition on the electric and magnetic charges of the fields. This condition comes from the requirement that the gauge transformation applied to $\Phi_1, A_\mu , C_\mu$ in \eqref{phiapprox} also leave the expanded Lagrangian ${\cal L}_{1+2}$ in \eqref{Lscalar6} invariant, especially with respect to the added field $\Phi_2$ which also couples to the same $A_\mu , C_\mu$ as $\Phi_1$. This mutual coupling of $\Phi_1 , \Phi_2$ to $A_\mu , C_\mu$ is the source of the proposed extra condition above. Applying \eqref{gauge-cond} to ${\cal L}_{1+2}$ in \eqref{Lscalar6} leads to
\begin{eqnarray}
\label{Lscalar7}
{\cal L}_{1+2} &=& \frac{1}{2} \partial_{\mu} \eta \partial^{\mu} \eta + \partial_{\mu}\Phi_2^*\partial^{\mu}\Phi_2 - \frac{1}{16 \pi}F_{\mu\nu}F^{\mu\nu}-\frac{1}{16 \pi}G_{\mu\nu}G^{\mu\nu}
+ m^2\eta^2(x) - \lambda v\eta^3(x) - \frac{1}{4}\lambda\eta^4(x) \nonumber \\
&+& \frac{1}{2}\left(v^2+2v\eta(x)+\eta^2(x)\right) \left( q_e^2A_{\mu}A^{\mu}+q_m^2C_{\mu}C^{\mu}+2q_e q_mA_{\mu}C^{\mu} \right) +J^{\mu}_2 \left( q'_e A_\mu + q'_m C_\mu \right)  \\
&+&\left((q '_e)^2 A_{\mu} A^\mu + (q'_m )^2 C_{\mu} C^\mu + 2 q'_e q'_m A_\mu C^\mu \right)\Phi_2^*\Phi_2 \nonumber ~.
\end{eqnarray}

Now in addition to the gauge transformation which we had to perform in the section IIA, we also performed a duality rotation as given in equations \eqref{AC-dual} \eqref{AC-inter2} which gave the Lagrangian in \eqref{Lscalar3}. In the final form of the Lagrangian ${\cal L}_1$, the remaining scalar field, $\eta$, only coupled to the magnetic gauge potential $C_\mu$ with a physical magnetic charge $g=\sqrt{q_e^2 + q_m ^2}$. Since the $A_\mu , C_\mu$ in this section are the same as in section IIA we must also perform the duality rotation from \eqref{AC-dual} to the new $\Phi_2$ terms in \eqref{Lscalar7} which can be written out as. 
\begin{eqnarray}
\label{AC-inter4}
&&\left(A_{\mu}^{\prime}, C_{\mu}^{\prime}\right)\left(
                                           \begin{array}{cc}
                                             \cos\theta & -\sin\theta \\
                                             \sin\theta & \cos\theta \\
                                           \end{array}
                                         \right)\left(
                                    \begin{array}{cc}
                                      (q'_e)^2 & q'_eq'_m \\
                                      q'_eq'_m & (q'_m)^2 \\
                                    \end{array}
                                  \right)\left(
                                           \begin{array}{cc}
                                             \cos\theta & \sin\theta \\
                                             -\sin\theta & \cos\theta \\
                                           \end{array}
                                         \right)
                                  \left(
                                           \begin{array}{c}
                                             A^{\mu \prime} \\
                                             C^{\mu \prime} \\
                                           \end{array}
                                         \right) \Phi_2^*\Phi_2 ~.
\end{eqnarray}
From section IIA $\theta$ is defined by $\cos\theta = \frac{q_m}{\sqrt{q_e^2+q_m^2}}$ and 
$\sin\theta = \frac{q_e}{\sqrt{q_e^2+q_m^2}}$. Performing the matrix multiplication gives 
\begin{eqnarray}
\label{AC-inter5}
&&\frac{\Phi_2^*\Phi_2}{q_e^2+ q_m^2} \left(A_{\mu}^{\prime}, C_{\mu}^{\prime}\right)
                                        \left(
                                    \begin{array}{cc}
                                      (q'_e q_m -q'_m q_e )^2 & ([q'_m]^2-[q'_e]^2)q_e q_m+(q_m^2-q_e^2)q'_eq'_m \\
                                      ([q'_m]^2-[q'_e]^2)q_e q_m+(q_m^2-q_e^2)q'_eq'_m & (q'_e q_m + q'_m q_e )^2 \\
                                    \end{array}
																		    \right)
                                        \left(
                                           \begin{array}{c}
                                             A^{\mu \prime} \\
                                             C^{\mu \prime} \\
                                           \end{array}
                                         \right)
\end{eqnarray}
In order to diagonalize matrix and get rid of the problematic cross terms, $A_{\mu}^{\prime} C^{\mu \prime}$, we require the following conditions
\begin{equation}
\label{dual-cond}
q_e = \pm q_m ~~~ {\rm and} ~~~ q_e ' = \mp q_m' ~.
\end{equation}
These are the second, non-Dirac conditions. As before, the reason for requiring these conditions is that both $\Phi _1 , \Phi _2$ couple to both $A_\mu , C_\mu$. Unlike the condition in \eqref{gauge-cond} which comes from the gauge transformation, these conditions arise from the duality transformation.  Actually once the condition, $q_e = \pm q_m$, is imposed the second condition 
({\it i.e.} $q_e ' = \mp q_m'$) follows from \eqref{gauge-cond}. Using the conditions \eqref{dual-cond} in \eqref{AC-inter5} leads to  
\begin{equation}
\label{AC-inter6}
\frac{\Phi_2^*\Phi_2}{q_e^2+ q_m^2} \left(A_{\mu}^{\prime}, C_{\mu}^{\prime}\right)
                                        \left(
                                    \begin{array}{cc}
                                      4 (q_e' q_e)^2 & 0 \\
                                      0 & 0 \\
                                    \end{array}
																		    \right)
                                        \left(
                                           \begin{array}{c}
                                             A^{\mu \prime} \\
                                             C^{\mu \prime} \\
                                           \end{array}
                                         \right) \rightarrow \frac{ 4 (q_e' q_e)^2}{q_e^2+ q_m^2} A'_\mu A^{' \mu} \Phi_2^*\Phi_2~.
\end{equation}
Next we examine what happens to the $J^{\mu}_2 (q'_e A_\mu + q'_m C_\mu )$ term in \eqref{Lscalar7} under the duality rotation. We find
\begin{equation}
\label{AC-inter7}
\left(q_e^{\prime}, q_m^{\prime}\right)
                                        \left(
                                           \begin{array}{cc}
                                             \cos\theta & \sin\theta \\
                                             -\sin\theta & \cos\theta \\
                                           \end{array}
                                         \right)
                                        \left(
                                           \begin{array}{c}
                                             A^{\mu \prime} \\
                                             C^{\mu \prime} \\
                                           \end{array}
                                         \right) \rightarrow \frac{1}{\sqrt{q_e^2+q_m^2}} \left( (q_e 'q_m - q_m' q_e) A^{\mu '} + (q_e 'q_m + q_m' q_e) C^{\mu '}\right) ~,
\end{equation}
where we have replaced the $\cos \theta$ and $\sin \theta$ by their values in terms of $q_e$ and $q_m$. Finally, applying the conditions from equation \eqref{dual-cond} we arrive at the dual rotated interaction between $J^{\mu}_2$ and $A_\mu '$
\begin{equation}
\label{J-AC}
J^{\mu}_2 (q'_e A_\mu + q'_m C_\mu ) \rightarrow \frac{2 q_e 'q_e}{\sqrt{q_e^2+q_m^2}} A^{\mu '} J^{\mu}_2  ~.
\end{equation} 
Collecting all of these results together from equations \eqref{CC-inter} \eqref{AC-inter6} \eqref{J-AC}, and using these in \eqref{Lscalar7} yields
\begin{eqnarray}
\label{Lscalar8}
{\cal L}_{1+2} &=& \frac{1}{2} \partial_{\mu} \eta \partial^{\mu} \eta + \partial_{\mu}\Phi_2^*\partial^{\mu}\Phi_2 - \frac{1}{16 \pi}F_{\mu\nu}F^{\mu\nu}-\frac{1}{16 \pi}G_{\mu\nu}G^{\mu\nu}
+ m^2\eta^2(x) - \lambda v\eta^3(x) - \frac{1}{4}\lambda\eta^4(x) \nonumber \\
&+& \left(\frac{1}{2}v^2+\frac{1}{2}v\eta(x) + \frac{1}{2}\eta^2(x)\right)\left(q_e^2+q_m^2\right)C_{\mu}C^{\mu} + \frac{2 q_e 'q_e}{\sqrt{q_e^2+q_m^2}} A^{\mu } J^{\mu}_2 
+\frac{ 4 (q_e' q_e)^2}{q_e^2+ q_m^2} A_\mu A^{ \mu} \Phi_2^*\Phi_2  ~.
\end{eqnarray} 
In this last step we have dropped the primes. From the last line of \eqref{Lscalar8} one can read off the physical magnetic and electric charges, $g$ and $e$, in terms of the initial Lagrangian parameters $q_e, q_m, q_e ' , q_m '$. These are
\begin{equation}
\label{eg-phys}
g = \sqrt{q_e ^2 + q_m ^2} ~~~ ; ~~~ e =\frac{2 q_e 'q_e}{\sqrt{q_e^2+q_m^2}} = \frac{2q' _e q_e}{g} 
\end{equation}

The point to notice about the final form of the Lagrangian ${\cal L}_{1+2}$ is that now $A_\mu$ is no longer decoupled, but couples directly to $\Phi_2$ via the last two terms in 
\eqref{Lscalar8}, and $C_\mu$ couples directly to $\eta$ -- both $C_\mu$ and $A_\mu$ interact non-trivially.

\section{Physical consequences of additional quantization conditions}

In the previous section we have shown how additional, non-Dirac quantization conditions arise between the initial electric/magnetic charges -- $q_e, q_m, q_e ' , q_m '$ -- of the fields 
$\Phi _1 , \Phi _2$ in the Cabbibo-Ferrari two-potential approach to magnetic charge. These non-Dirac conditions come from requiring that some combination of gauge transformations and duality rotations 
leads to a consistent, gauge invariant theory. We repeat these additional condition here for convenience  
\begin{equation}
\label{gauge-dual}
\frac{q_e'}{q_e}=-\frac{q_m'}{q_m} ~~~{\rm and}~~~ q_e = \pm q_m  ~.
\end{equation}
In \eqref{gauge-cond} there is an additional condition, $q_e ' = \mp q_m'$, which we leave off here since it is redundant -- it comes from combining the two conditions in \eqref{gauge-dual}. 

The first condition can be re-written as $\frac{q_e'}{q_m '}=-\frac{q_e}{q_m}$ so that the ratio of initial electric to magnetic charges of the two scalar fields are not equal {\it i.e.} $\frac{q_e'}{q_m '} \ne \frac{q_e}{q_m}$. This result can be connected with the well known statement ``If particles/fields all have the same ratio of magnetic to electric charge then we can make a duality rotation by choosing $\theta$ so that $\rho _m=0$ and ${\bf J}_m =0$" \cite{jackson}. Therefore, the first condition in \eqref{gauge-dual}, which results from the gauge transformation, ensures that one has a non-trivial theory with the final Lagrangian in \eqref{Lscalar8} having fields with both electric and magnetic charges as given in \eqref{eg-phys}. 

If one looks at the last equation in \eqref{eg-phys} one sees that it can be re-written as $e g = 2 q_e q_e '$ so that the physical electric and magnetic charges, $e$ and $g$, are given in terms of only the starting electric charges, $q_e$ and $q_e '$, of the two scalar fields. The general conclusion is that the new, non-Dirac conditions are a re-parametrization of the physical charges in terms of the original Lagrangian charges -- $q_e, q_e ', q_m, q_m '$. Directly from \eqref{eg-phys} one can see that the physical electric and magnetic charges are given by some combination of the initial ``electric" and ``magnetic" parameters. 

Next we look into what happens with the usual Dirac quantization condition $eg = n \frac{\hbar}{2}$ in the above Cabbibo-Ferrari approach with symmetry breaking. In the two-potential approach there is no Dirac string, no need for the Dirac veto and thus one might expect there will be no Dirac quantization condition. However, the Dirac quantization condition can be obtained independently
of the Dirac string argument by requiring that the field angular momentum of the electric-magnetic charge system be quantization in half-integer units of $\hbar$ \cite{saha}. Moreover, since we have hidden the magnetic gauge symmetry associated with $C_\mu$ via symmetry breaking, the magnetic field produced by the magnetic charge will have a Yukawa character and this will change the quantization condition.

\begin{figure}
  \centering
	\includegraphics[trim = 0mm 0mm 0mm 0mm, clip, width=9.0cm]{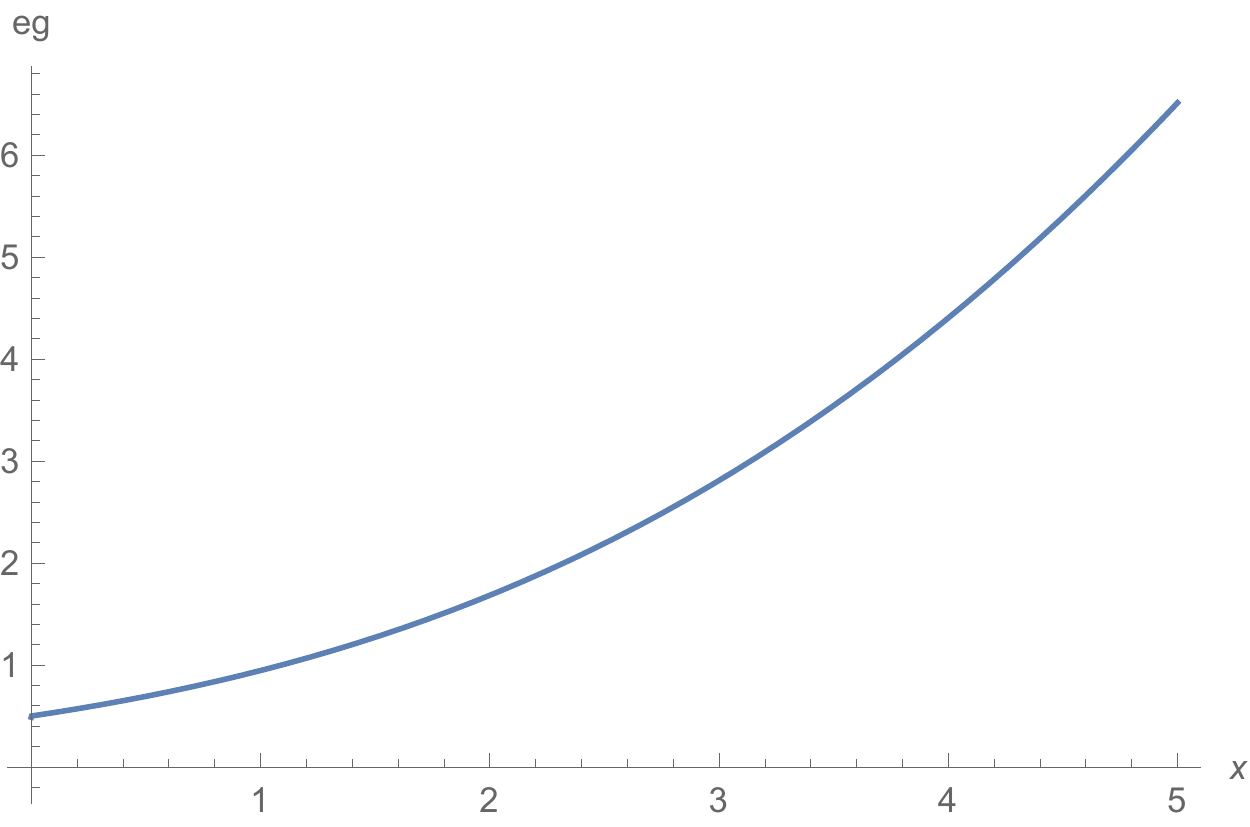}
\caption{\it The plot of $eg$ versus $x = m_C R$ for the case $n=1$. In the limit $m_C \to 0$ ({\it i.e.} $x \to 0$) one recovers $eg =\frac{1}{2}$. For any non-zero mass and non-zero separation  $x>0$ one finds that the product $eg$ will be greater than for the usual Dirac condition. This implies that the strength of the magnetic charge is even larger in the presence of symmetry breaking.}
\label{fig1}
\end{figure}

We begin by placing the magnetic charge associated with $\eta$ at the origin and the electric charge associated with 
$\Phi_2$ a distance ${\bf R}$ from $\eta$.  The respective four-vector potentials produced by this configuration are 
\begin{equation}
\label{egpot}
A ^{\mu} = \left( \frac{e}{r'} , 0, 0, 0 \right)  ~~~ {\rm and}~~~
C ^{\mu} = \left( \frac{g e^{-m_C r}}{r} , 0, 0, 0 \right)
\end{equation}
where $r' = \vert {\bf r} - {\bf R} \vert$. The field/particle $\Phi_2$ produces a Coulomb potential, while
the particle/field $\eta$ produces a Yukawa potential. Using the definitions of the electric and magnetic fields from 
\eqref{newEB-pot} gives
 \begin{equation}
\label{EB-yukawa}
{\bf E}=-\nabla\phi_{e} =  - \frac{e {\bf r'}}{{r'} ^3} ~~~~;~~~~
{\bf B}=-\nabla\phi_{m} = -  \frac{g e^{-m_C r}}{r^2} \left( \frac{1}{r} + m_C \right) {\bf r}~.
\end{equation} 
The field angular momentum associated with the electric and magnetic fields in \eqref{EB-yukawa} is given by 
\begin{equation}
\label{ang3d1}
{\bf L} = \frac{1}{4 \pi} \int {\bf r} \times ({\bf E} \times {\bf B}) d^3 x ~.
\end{equation}
Since the fields are radial we will, without loss of generality, place the electrically charged particle, $\Phi _2$, along the z-axis
{\it i.e.} ${\bf R} = R {\bf {\hat z}}$. The field angular momentum in \eqref{ang3d1} for this set up only has a non-vanishing
component in the z-direction. After performing the $d \varphi$ integration one finds
\begin{equation}
\label{angfin}
{\bf L} = - \frac{egR}{2} \int _0 ^{\infty} r^2 dr \left[ \int _{-1} ^1
d(cos \theta) \frac{e^{-m_C r}}{(r^2 + R^2 - 2rR cos \theta) ^{3/2}}
\Big( m_C + \frac{1}{r} \Big) \big( 1 - cos ^2 \theta \big)  \right] {\bf {\hat z}} ~.
\end{equation} 
Doing the remaining $r, \theta$ integrations \cite{singleton2} \cite{chatterjee} \cite{chatterjee2}  gives 
\begin{equation}
\label{angfin1}
L_z = |{\bf L}|=  \frac{2 eg}{m_C ^2 R^2} \Big[ 1 - (1+m_C R)e^{-m_C R} \Big] ~.
\end{equation}
In the $m_C \rightarrow 0$ limit we recover the results found in \cite{saha} (also see \cite{cart} for a more accessible and modern
derivation of this result for the $m_C=0$ case). Now making the assumption that $L_z = n\frac{\hbar}{2}$ and defining $x= m_C R$ we 
can solve \eqref{angfin1} for the product of charges $eg$ to find  
\begin{equation}
\label{angfin1a}
eg = \frac{n \hbar}{4} \left( \frac{x^2}{ 1 - (1+x)e^{-x}} \right)  \rightarrow \frac{1}{4} \left( \frac{x^2}{ 1 - (1+x)e^{-x}} \right)~.
\end{equation}
In the last limit we have taken the lowest option of $n=1$ and set $\hbar = 1$. The $x$-dependent factor in parenthesis represents 
the deviations, due to symmetry breaking, from the usual Dirac condition. The plot of \eqref{angfin1a} is shown
in figure \eqref{fig1}. From this one can see that when we turn off symmetry break ({\it i.e.} $m_C$ and $x \to 0$) we recover 
the expected result of $eg =\frac{1}{2}$ for the $n=1$ case given in the graph. When we turn on symmetry break and $m_C >0$ we see
that $eg > \frac{1}{2}$. This implies that the strength of the magnetic coupling will be stronger in the presence of symmetry breaking 
as compared to when there is no symmetry breaking. This makes sense in the following way: due to the presence of a mass for $C_\mu$ the
magnetic field will fall off more rapidly than in the Coulomb case as can be seen \eqref{EB-yukawa}. To compensate for this weaker magnetic field, the magnetic coupling $g$ must be larger in order to still give an angular momentum of $L_z =\frac{n \hbar}{2} \to \frac{1}{2}$.  

The arguments above indicate that the magnetic coupling, determined by the magnetic fine structure constant $\alpha _g = \frac{g^2}{\hbar c}$, will be non-perturbatively large. This is in contrast with the electric fine structure constant, $\alpha _e = \frac{e^2}{\hbar c} \approx \frac{1}{137}$, which is perturbatively small. The large value of $\alpha_g$ strongly indicates the $U(1)$ symmetry connected with the magnetic charge will undergo a phase transition so that magnetic charge will be confined \cite{wilson} \cite{schwing} \cite{guth}. In references \cite{tousi} lattice gauge calculations indicated that the critical value of the coupling for a $U(1)$ theory to become confining is $\alpha _c \sim {\cal O} (1)$. For the present case $\alpha _g \approx 137/4$ and this is clearly well above unity so that magnetic charge in this model should be confined. Further one could propose that it is this non-pertubative value of $\alpha _g$ which drives the mass term for the magnetic photon, as in the old techni-color models (for a recent review see \cite{TC-review}). In this picture a monopole-antimonopole pair would form a condensate due to the strong coupling implied by $\alpha _g$, and this condensate would mix with the magnetic gauge boson to generate a mass, exactly as strongly interacting techni-quarks were supposed to form a condensate which would mix with the $W^\pm$ and $Z$ bosons to give them their masses.

\section{Conclusions}

We have examined the Cabibbo-Ferrari, two-potential approach to magnetic charge with two complex scalar fields, each carrying different
initial ``electric" and ``magnetic" couplings -- $q_e, q_e ', q_m, q_m '$. In this formulation the Dirac string is replaced by 
an additional gauge potential, $C_\mu$. In the usual Dirac string approach the singular string has to be ``hidden" or made non-observable. 
This is accomplished by imposing the Dirac condition on the electric and magnetic charges, namely $eg = n\frac{\hbar}{2}$. Here we need to
``hide" the extra gauge potential. This is accomplished by assuming that $\Phi_1$ has a Higgs-like self-interaction which breaks the magnetic,
$U(1)$ symmetry and makes $C_\mu$ massive so that the magnetic vector potential is un-observable, at least up to an energy scale of 
$m_C c^2$.  With only a single scalar field, $\Phi_1$, we find that we have a trivial system in the sense that the massless, potential, $A_\mu$, which decouples from everything else, and there is only one matter field, $\eta$, which carries only a magnetic charge. We next introduced a second scalar field, $\Phi_2$, with different initial ``electric" and ``magnetic" charges. With this additional matter field we found that one needed to impose non-Dirac conditions on the Lagrangian charges, $q_e, q_m, q_e', q_m'$ in order for the Lagrangian to be properly gauge invariant and in order for the duality transformation to give interaction terms that avoided problematic mixed terms proportional to $A_\mu C^\mu$. These additional, non-Dirac conditions were given in equations \eqref{gauge-cond} and \eqref{dual-cond}. 

We also found that the magnetic photon gaining a mass modified the usual Dirac condition since the magnetic field now had a Yukakwa behavior which changed the field angular momentum of electric cahrge-magnetic charge system to \eqref{angfin1}. This modified Dirac quantization condition is given in \eqref{angfin1a}. We found that this modified condition implied a larger magnetic charge as compared to the usual Dirac case. In addition, this new Dirac-like quantization condition involved not only the charges, $e$ and $g$, but also the parameter $x = m_C R$. This is expected since with symmetry breaking one introduces a mass/distance scale ({\it i.e.} $m_C$ or $d=1/m_C$) which should then show up in the quantization condition. An interesting conjecture is that the Dirac-like quantization condition in the presence of symmetry breaking might lead to a mass quantization condition. This idea of mass quantization coming from Dirac electric/magnetic charge quantization in the presence of symmetry breaking will be investigated in a future work.        

\section*{Acknowledgements}

This work was supported by Grant $\Phi.0755$  in fundamental research in natural sciences by the Ministry of Education and Science of Republic of Kazakhstan.

\end{document}